\documentclass[aps,prl,preprint,shopacs,groupedaddress]{revtex4-1}
\usepackage{graphicx}
\usepackage{amsmath}

\begin{document}

\preprint{SAND-2013-8384J}

\title{Theory of melting at high pressures: Amending Density Functional Theory with Quantum Monte Carlo}


\author{L. Shulenburger}
\email[]{lshulen@sandia.gov}
\author{M. P. Desjarlais}
\email[]{mpdesja@sandia.gov}
\author{T. R. Mattsson}
\email[]{trmatts@sandia.gov}
\affiliation{Sandia National Laboratories, Albuquerque, New Mexico 87185, USA}


\date{\today}

\begin{abstract}
We present an improved first-principles description of melting under pressure based on thermodynamic integration comparing Density Functional Theory (DFT) and quantum Monte Carlo (QMC) treatments of the system.  The method is applied to address the longstanding discrepancy between density functional theory (DFT) calculations and diamond anvil cell (DAC) experiments on the melting curve of xenon, a noble gas solid where van der Waals binding is challenging for traditional DFT methods. The calculations show excellent agreement with data below 20 GPa and that the high-pressure melt curve is well described by a Lindemann behavior up to at least 80 GPa, a finding in stark contrast to DAC data.
\end{abstract}



\maketitle


The high pressure melt line of simple materials carries great significance in both
purely theoretical and in practical applications.  For instance, the rapid decrease followed by
suspected increase in the melting temperature of lithium under pressure is a bellwether for the complex series of
solid phases that exist at lower temperatures.\cite{lithium-melt} Furthermore, the
onset of melt triggers a dramatic loss of mechanical strength of a material, with significant changes in dynamic behavior following.
In fact, the point where a material melts under shock compression is one of the key properties that can
distinguish between possible scenarios for planetary accretion.\cite{turbulent-mixing-planetary-accretion} 
Although diamond anvil cell (DAC) experiments remain the most versatile experimental technique
for probing high pressure melting behavior, they have also been a source of controversy.
Important examples exist in the literature of melt lines showing an anomalous change in slope
under pressure that were contradicted by either shock experiments or later DAC experiments.\cite{dewale-ta-melt,dewale-Fe-melt}
An as yet unchallenged melt line of this type is exhibited by xenon and other noble gases - which are of particular importance
due to their inert nature. The high pressure behavior of the noble gases is a fundamental test of the DAC methodology and as such deserves special scrutiny. In this letter, we specifically
consider the behavior of xenon and find that the high-pressure melt curve is well described by a traditional melting curve.

As alluded to above, the experimentally obtained melting
curve for xenon exhibits an interesting feature when probed in the diamond anvil cell, abruptly
flattening out at pressures above 25 GPa.\cite{dac-xe-melt}. The observation prompted much theoretical attention,
including applying quantum mechanical simulation techniques to the problem.\cite{belonoshko-xe-melt}
These techniques, lead by density functional theory (DFT), are uniquely suited
to the study of extreme conditions as their fundamental approximations are not affected
by the presence of temperature or pressure, if a calculation is accurate near ambient
conditions, the method is also likely to be accurate at high pressure.  DFT applied
to xenon finds a Lindemann like melt curve in contrast to the experiments.\cite{belonoshko-xe-melt}

The accuracy of DFT calculations of noble gases, however, is not to be taken for granted since fundamental uncertainties
remain regarding calculations of systems where van der Waals interactions are significant.
Standard semi-local functionals such as the local density approximation (LDA) tend to over-bind the noble gases due to a spurious self 
interaction of the electrons in regions of low density.  Improved functionals 
such as AM05\cite{AM05} remove this self-interaction, but as a result do not 
bind noble gas solids at all. Despite much progress in the area of dispersion corrected density functional theory \cite{vdw-dft},
cases involving the transition where dispersion dominated bonding gives way to covalent-or metallic bonding remains a challenge.
Xenon presents a canonical example of this effect and as a result its behavior is greatly affected by pressure.  Xenon turns metallic under 
moderate shock compression\cite{nellis-shock-xe} and although xenon is a narrow-range cryogenic 
liquid at normal pressure with melting and boiling points of 161.4 K and 165.0 K, 
respectively, the melting point at 20 GPa is above 2500 K. 

These significant theoretical challenges necessitate the application of a complementary technique whose 
approximations are not tied to the local behavior of the electrons.  A promising candidate
from this point of view is diffusion quantum Monte Carlo (DMC).\cite{dmc-review}
Whereas the approximation made in DFT calculations requires the consideration of an 
effective Hamiltonian, DMC treats the Hamiltonian exactly.  Therefore,
DMC can accurately study van der Waals interactions and has been 
successfully applied to lighter noble gas solids \cite{cambridge-dmc-noble-gas} 
and the interactions between filled shell molecules.\cite{casula-benzene-dimer,beaudet-h2-benzene}

In order to thoroughly investigate the performance of DMC for xenon, we focused on
the three fundamental approximations that would be necessary in the calculations.
These approximations are the pseudopotential
approximation that is necessary for computational efficiency, the fixed node approximation which is
necessary to mitigate the fermion sign problem, and the finite size approximation where calculations on
modest sized supercells are used to determine properties xenon in the thermodynamic limit.

As a test of these approximations, the energy versus volume for the FCC crystal is used as
a benchmark.  Calculations of a 32 atom supercell, using the finite size correction methods
employed in the rest of the paper with two different starting points are considered.  Firstly 
pseudopotentials and nodal surfaces from the LDA are used as input to the DMC calculations.  Then
the processes is repeated with pseudopotentials and nodal surfaces from AM05, allowing a
sensitivity test to the form of these approximations.

\begin{figure}
\includegraphics[width=4.0in,angle=-90]{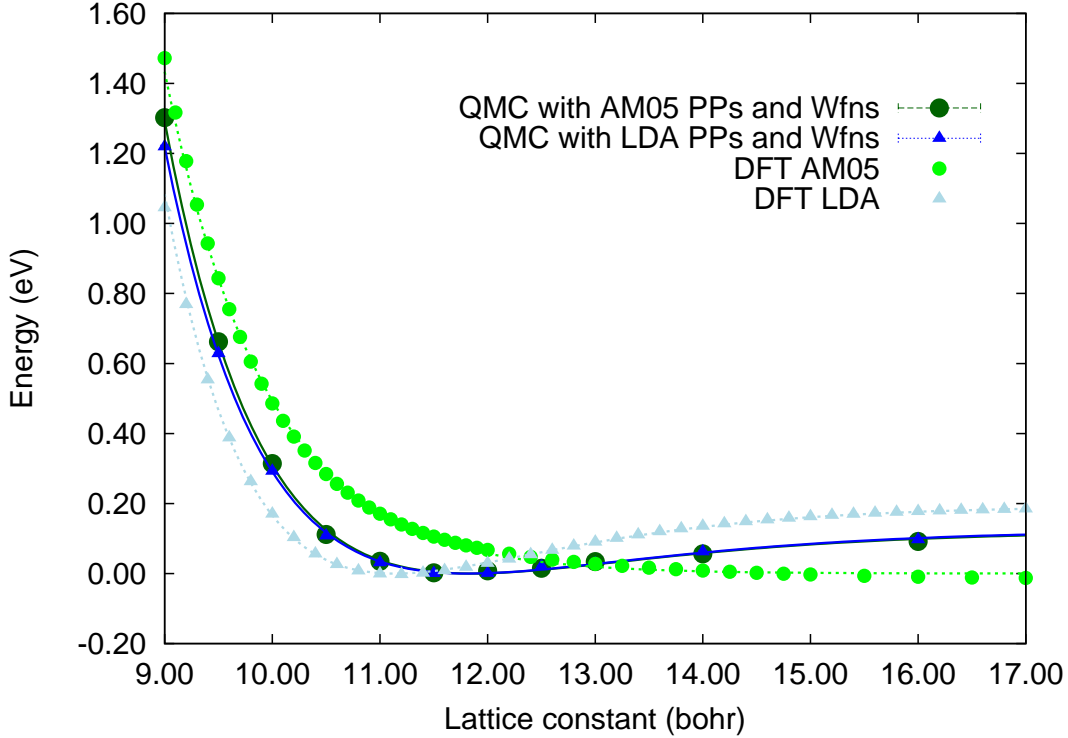}
\caption{(color online) Energy of a unit cell of FCC xenon calculated with DFT and DMC.  The
dotted lines correspond to Vinet fits to the DFT calculations.  The solid lines
correspond to Vinet fits to the DMC calculations.  The triangles correspond to
DFT or DMC simulations based on the LDA and the circles DFT or DMC based on AM05.
\label{cold-curves}}
\end{figure}

The results of this test are shown in Fig\,\ref{cold-curves}.  We find that the DMC 
results are independent of the trial wavefunctions and pseudopotentials to the level required for
this work. Fitting the DMC energy versus volume curve with a Vinet form\cite{vinet} gives 
a lattice constant varying by only $0.25\% \pm 0.61\%$ when changing from LDA 
to AM05 trial wavefunctions and a bulk modulus varying by only $0.4\% \pm 0.8\%$.  
For this reason we conclude the errors arising from nodal and pseudopotential approximations are small 
for these DMC calculations of xenon.

Despite this evidence that DMC is ideally suited for the calculation of the properties
of xenon under pressure, one important wrinkle remains.  Direct calculations of melting are not 
currently feasible with DMC for anything beyond the lightest of elements.  Fortunately,
a solution to this problem has recently been proposed: thermodynamic integration can 
be used to connect the accuracy of the DMC calculations with the speed and efficiency
of DFT based molecular dynamics.\cite{sola-alfe-fe-melt}  Using this technique, Sola and
Alf\'e found that DMC calculations favored the solid phase in calculations of the melting
of iron under pressure.  This result was in disagreement with DAC
experiments.\cite{dac-melt-fe}  A potential concern with this result is that QMC methods (both VMC
and DMC) being variational tend to produce relatively lower total energies for more ordered states 
(in this case solids versus liquids).  This effect is because the trial wavefunctions
used tend to be rather simple compared to the true many body wavefunctions
and typically do not increase in complexity for the less ordered phases.
Thus simpler phases where the wavefunction is closer to the many body wavefunction 
tend to have a smaller positive fixed node error than that for a more complex
phase.

In light of this and because the approach is new, we elected to null-test
the method by calculating the melting temperature of aluminum at 120 GPa.  
This material and condition were chosen because shock experiments, diamond
anvil cell experiments and DFT calculations all agree as to the melting
temperature\cite{al-melt-paper}. If the QMC free energies were
biased towards the solid phase then the melting temperature would be
overestimated using this method.
Relative energies between the snapshots of the same phase for aluminum agreed very 
well between the DMC and DFT, giving confidence that the DFT dynamics
were close to the DMC ones.  Additionally, the shift in free energy between
the solid and liquid was very small, $0.202 \pm 0.100$ meV/atom, leading
to a temperature shift of only $2.3 \pm 1.2$ K.  This result is well within
the errors of the method and experimental accuracy for melting under pressure.
Furthermore, this test shows that the thermodynamic integration method does not suffer from notable
systematic errors when the DMC is performed with a relatively simple
trial wavefunction.

In applying this approach to the melting of xenon we start by calculating the melting line at
two points using DFT based molecular dynamics.  Specifically following the work of 
Root et al.\cite{root-xe-hugoniot} we performed calculations using VASP\cite{vasp} within
the AM05\cite{AM05} density functional.  We used two-phase coexistence simulations to establish 
the relative free energies between the solid FCC and liquid phases of the xenon at high pressure.  
Two densities were selected for these simulations, 7.27 g/cc and 10.0 g/cc.  These simulations were 
performed in both the NVE and NVT ensembles, using the consistency between the two to 
check that the technical parameters of the simulations were converged.  Indeed, we found that 
for the higher density simulation, calculations with 214 xenon atoms found a melt temperature 
of 6000 K in the NVT ensemble, but the NVE yielded a lower value.  This suggested that 
larger simulation cells were necessary and upon consideration of cells doubled in size in the 
direction perpendicular to the interface (428 atoms) the results agreed, yielding two points 
at which the Gibbs free energy of the two phases were equal: 24.4 GPa and 3000 K for 7.27 g/cc 
and 74.4 GPa and 5600 K for 10.0 g/cc.  

\begin{figure}
\includegraphics[width=4.0in,angle=-90]{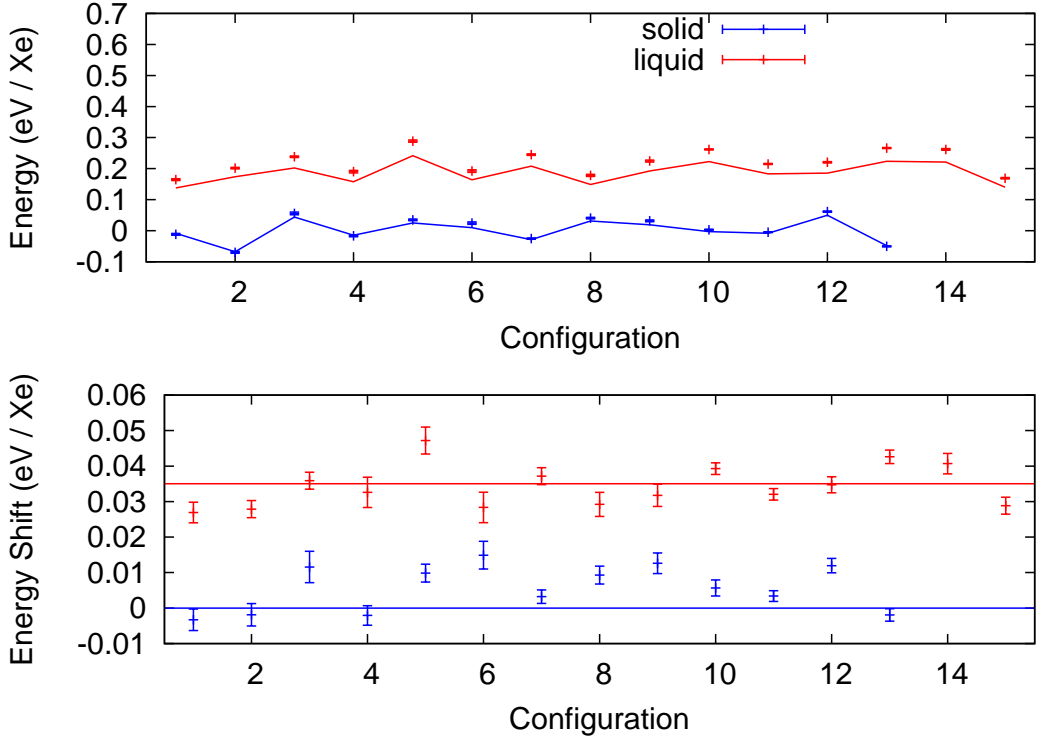}
\caption{(color online) Top panel: DMC energies corresponding to configurations representative
of solid (blue triangles) and liquid (red squares) xenon, generated with QMD on 
108 atom systems.  The solid lines connect DFT energies calculated on the same
configurations.  An independent offset is added to the DMC and DFT calculations so that the
average energy of the solid snapshots in each method is 0.  Bottom panel: DMC-DFT
energy differences for the same configurations.  The average DMC-DFT energy difference
for the solid is subtracted from all points.  Lines represent the average of the
energy differences between DMC and DFT in the solid and the liquid.
\label{xe-snapshot-en}}
\end{figure}

From this foundation, we followed Sola and Alf\'e \cite{sola-alfe-fe-melt} adding refinements to the
methodology to further reduce the uncertainty.
The change in free energy of a phase at a given temperature and pressure is calculated by taking snapshots 
from long DFT based molecular dynamics simulations and comparing the energy of those snapshots to
energies from DMC calculations.  Using this information, the change in the Helmholtz free
energy of each phase is found using a perturbation series of cumulants in the energy difference as:
\begin{equation}
\Delta F = \sum_{n=0}^{\infty} \frac{(1/k_B T)^{n-1}}{n!} \kappa_n
\label{thermo-int-eqn}
\end{equation}
where the $\kappa_n$'s are cumulants of the difference in internal energy between the DMC and DFT ensembles:
\begin{align}
\kappa_0 &= \left< \Delta U \right>_{\lambda=0} \nonumber \\
\kappa_1 &= \left< \Delta U^2 \right>_{\lambda=0} - \left< \Delta U \right>^2_{\lambda=0} \nonumber \\
& \quad \vdots
\label{cumulants-eqn}
\end{align}
or directly in terms of the partition function
\begin{equation}
\Delta F = -k_B T \left< e^{-\Delta U/k_B T} \right>_{\lambda=0}
\label{thermo-int-eqn2}
\end{equation}

where $\Delta U = U_{DMC}-U_{DFT}$ with $U_{DMC}$ and $U_{DFT}$ the potential energies of the DMC 
and DFT systems respectively and $\left<\right>_\lambda$ represents the thermal average in the 
ensemble generated by the potential energy function $U(\lambda) = \lambda U_{DMC} + (1-\lambda)U_{DFT}$.
  The approximation above is valid when $U_{DMC}$ and $U_{DFT}$ are sufficiently close so that 
the averages over all of state space can be approximated using a few configurations sampled 
from the ensemble of the reference system.  A necessary condition for this to be valid is that 
the higher order terms in Eq.\,\ref{thermo-int-eqn} are small and that the two approximations 
in Eq.~\ref{thermo-int-eqn} and Eq.~\ref{thermo-int-eqn2} yield very similar answers.  An example
of this methodology is found in Fig.~\ref{xe-snapshot-en}.  From this figure, it is
apparent that the total energies track each other well, again 
suggesting that DFT provides a faithful sampling of the energy landscape.  Quantitatively, 
Eq.\,\ref{thermo-int-eqn} bears this out, with the second term in the cumulant expansion 
being $1.5\%$ of the first one for the solid at 7.27 g/cc and $1.4\%$ for the liquid.  The bottom panel in Fig.~\ref{xe-snapshot-en} shows
the differences between the solid and the liquid snapshots after the average DMC-DFT
energy difference for the solid is subtracted for all points.  This 
shows visually that the DMC energy is on average $35.0$ meV/atom 
larger for the liquid snapshots than the corresponding DFT.

Once the change in the Helmholtz free energy is calculated, the change to
the melting temperature produced by DFT can be found using the formula
\begin{equation}
\Delta T^m \simeq \frac{G^{ls}}{S_{DFT}^{ls}}
\label{melt-shift-equation}
\end{equation}
where the superscript $ls$ indicates differences between liquid and solid,
$S_{DFT}^{ls}$ is the DFT entropy of melting.
The difference in the Gibbs free energy is $\Delta G \simeq \Delta F -
V\Delta p^2/2 B_T$ with $B_T$ the isothermal bulk modulus and $\Delta p$ the
change in pressure as the potential energy is changed from $U_{DFT}$ to $U_{DMC}$ at constant volume.
In the work of Sola and Alf\'e,\cite{sola-alfe-fe-melt} the corrections
to the Gibbs free energy are found to be small so that the value of $\Delta F$ at
constant V is also representative of $\Delta G$ at constant p.

Uncertainties in the size of the approximations made in this approach may be removed
by making a modification to the procedure.  Instead of performing a one shot calculation
of free energy at a single point in (V, T) space, an entire isotherm can be evaluated.  First, 
QMD calculations are performed at several different densities along the 3000K and 5600K
isotherms centered around the melt densities calculated with the two phase calculations.
Using the relation at constant temperature that
\begin{equation}
dF = - \int_{V_i}^{V_f} P dV + C {\rm ,}
\label{isothermal-thermodynamic-integration}
\end{equation}
relative Helmholtz free energies in each phase may be found.  The two phase calculation
allows for the relative free energies between these phases to be set using the Gibbs construction.

For the lower density case, these free energy curves were augmented by a shift in the relative
free energies using the 30.1 meV per Xe found with the above techniques.  Assuming that this 
free energy shift will be constant as a function of volume, the change in the melt line can be found
in two different ways. First, the change in the melt temperature at constant pressure is found
using Eq.~\ref{melt-shift-equation} since the additional thermodynamic information contained in the relative
free energy in each phase allows the isothermal bulk modulus and the change in entropy upon melt
to be calculated directly, yielding 82 GPa and 0.787 $k_B$ respectively.  This assumption of a rigid shift
in the free energy renders the second term in the change in the Gibbs free energy 0 because of a  zero
shift in pressure from one theory to the next.  Putting this all together, gives a shift in the melt temperature
to 3440 K at 24.4 GPa.  Second, one can use a Gibbs construction on the relative Helmholtz free energies and find
a pressure shift to a melt of 18.66 GPa at 3000 K.

\begin{figure}
\includegraphics[width=4.0in,angle=-90]{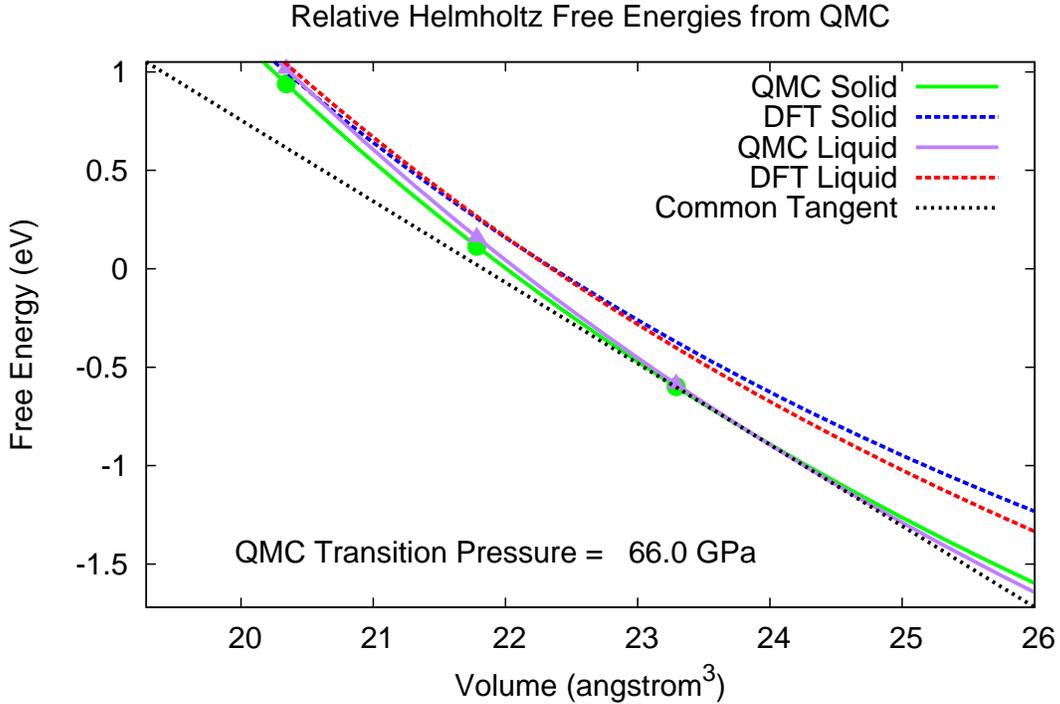}
\caption{(color online) Relative Helmholtz free energy of the solid and liquid phases at 5600K as determined by DFT using two phase calculations to establish the melt pressure and thermodynamic integration to find the relative free energies.  A common tangent to the QMC curves is also shown, establishing a new melt pressure of 66 GPa.
\label{free-energy-figure}}
\end{figure}

Finally, and most importantly, we take into account the effect of a change in pressure on the free energy differences of the melt near
10 g/cc where the discrepancy between theory and experiment is the largest. We here use thermodynamic integration at three different densities, allowing for information about changes in the size of the shift as a function of pressure to be considered, a notable effect in compressible
materials like xenon.
Doing so, we found that the pressure changes by 9 GPa upon switching from a DFT to a QMC ensemble while the isothermal bulk
modulus increases to 215 GPa.  These results are shown in Fig.~\ref{free-energy-figure}, which shows how the
relative free energies of the solid and liquid are changed by the thermodynamic integration.  Now the full change in 
the Gibbs free energy for each phase can be found, yielding a melting temperature of 5810 K at 74.4 GPa.  Had the
relative change in free energy from the thermodynamic integration been assumed to be constant, this would have yielded a higher melting temperature of 6130 K at 74.4 GPa.  Also, a pressure shift can be found as above, yielding a melting pressure of 66 GPa at 5600 K.

\begin{figure}
\includegraphics[width=4.0in,angle=-90]{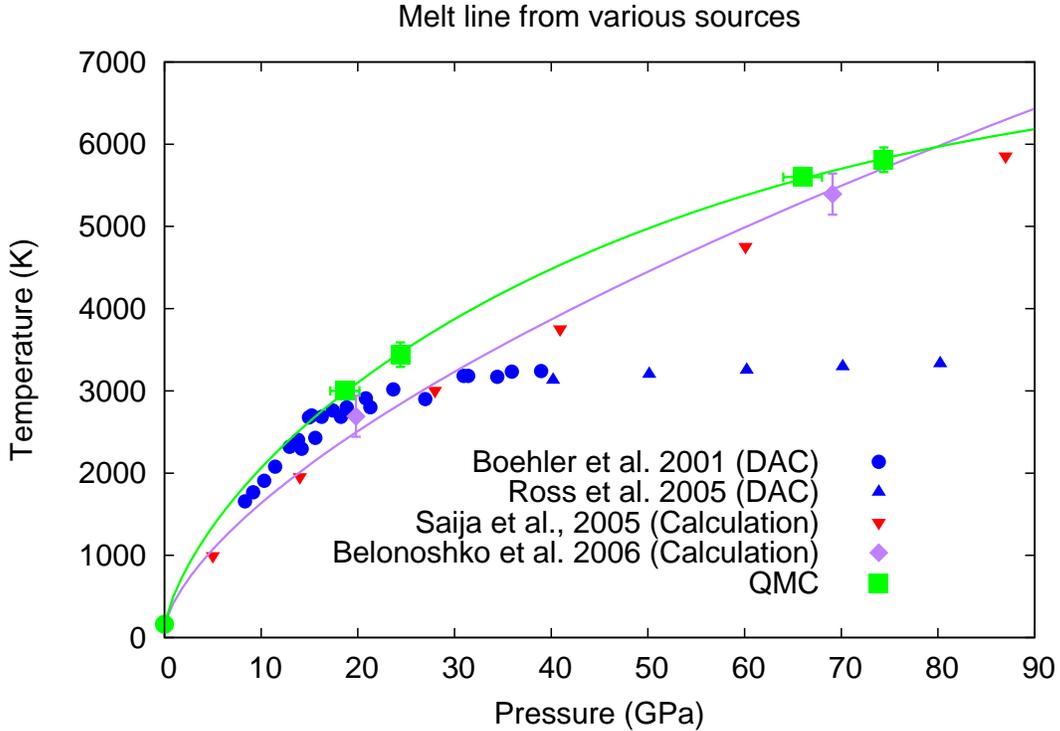}
\caption{(color online) Melting temperature of xenon as a function of
pressure obtained with various theoretical\cite{saija,belonoshko-xe-melt} and experimental\cite{boehler-xe-melt,dac-xe-melt}
techniques.  The horizontal error bars on the QMC points come from propagating the statistical uncertainty in the pressure shift technique, the vertical error bars from the temperature shift technique.  The quantum Monte Carlo data is fit with a Kechin form:
$T(P) = a(1+\frac{P}{b})^c e^{-dP})$\cite{ketchin-melt}.
\label{xe-melt-figure}}
\end{figure}

Taking these two points together with the
well established melting temperature at ambient pressure results in a
melt line shown in Fig.\,\ref{xe-melt-figure}.  The net effect
is to increase the disagreement between the high pressure melt line and
the DAC experiments.  A notional Kechin melt curve fitting these two 
high pressure points and the ambient pressure melting is shown in the figure.
The melting has been brought into better agreement
with the DAC at low pressures, but suggests that the flattening of the
melt curve at high pressures is not correct.  These results for xenon
suggest that the high pressure DAC experiments should be reexamined to rule
out either surface effects or non-hydrostatic stresses as the cause of the
flat melt line.\cite{explain-melt-anomalies}  This result might be achieved by exploiting a bulk probe
of the xenon structure such as x-ray diffraction rather than the speckle
field technique that was previously used.\cite{dac-xe-melt}

In addition to this result on xenon, we have provided validation of the thermodynamic integration approach
to using DMC\cite{sola-alfe-fe-melt} to inform high pressure melt boundaries by 
performing a test of the method on aluminum.  In the process we extended the methodology, improving
the accuracy for compressible materials.  This high-accuracy procedure can be
used to further explore the melting behavior of a wide variety of materials, thereby
contributing to the ability of hydrodynamic simulations to predictively model a wide
range of phenomena from inertial confinement fusion to planetary science.

The authors acknowledge helpful discussions with Ken Esler who provided
help in developing the xenon pseudopotentials and thank John Carpenter for the suggestion
to validate the method against the aluminum  melt line and checking the low pressure xenon behavior.
The calculations reported here were performed using the Sandia National Laboratories TLCC 
machines as well as the ACES cielo. The work was supported by the NNSA Science Campaigns and LNS was supported through the Predictive Theory and Modeling for Materials and Chemical Science program by the Basic Energy Science (BES), Department of Energy (DOE).
Sandia National Laboratories is a 
multiprogram laboratory managed and operated by Sandia Corporation, a wholly 
owned subsidiary of Lockheed Martin Corporation, for the U.S. Department of 
Energy's National Nuclear Security Administration under Contract No. DE-AC04-94AL85000.


\bibliography{xe-melt-bib}

\begin{thebibliography}{23}%
\makeatletter
\providecommand \@ifxundefined [1]{%
 \@ifx{#1\undefined}
}%
\providecommand \@ifnum [1]{%
 \ifnum #1\expandafter \@firstoftwo
 \else \expandafter \@secondoftwo
 \fi
}%
\providecommand \@ifx [1]{%
 \ifx #1\expandafter \@firstoftwo
 \else \expandafter \@secondoftwo
 \fi
}%
\providecommand \natexlab [1]{#1}%
\providecommand \enquote  [1]{``#1''}%
\providecommand \bibnamefont  [1]{#1}%
\providecommand \bibfnamefont [1]{#1}%
\providecommand \citenamefont [1]{#1}%
\providecommand \href@noop [0]{\@secondoftwo}%
\providecommand \href [0]{\begingroup \@sanitize@url \@href}%
\providecommand \@href[1]{\@@startlink{#1}\@@href}%
\providecommand \@@href[1]{\endgroup#1\@@endlink}%
\providecommand \@sanitize@url [0]{\catcode `\\12\catcode `\$12\catcode
  `\&12\catcode `\#12\catcode `\^12\catcode `\_12\catcode `\%12\relax}%
\providecommand \@@startlink[1]{}%
\providecommand \@@endlink[0]{}%
\providecommand \url  [0]{\begingroup\@sanitize@url \@url }%
\providecommand \@url [1]{\endgroup\@href {#1}{\urlprefix }}%
\providecommand \urlprefix  [0]{URL }%
\providecommand \Eprint [0]{\href }%
\providecommand \doibase [0]{http://dx.doi.org/}%
\providecommand \selectlanguage [0]{\@gobble}%
\providecommand \bibinfo  [0]{\@secondoftwo}%
\providecommand \bibfield  [0]{\@secondoftwo}%
\providecommand \translation [1]{[#1]}%
\providecommand \BibitemOpen [0]{}%
\providecommand \bibitemStop [0]{}%
\providecommand \bibitemNoStop [0]{.\EOS\space}%
\providecommand \EOS [0]{\spacefactor3000\relax}%
\providecommand \BibitemShut  [1]{\csname bibitem#1\endcsname}%
\let\auto@bib@innerbib\@empty
\bibitem [{\citenamefont {Guillaume}\ \emph {et~al.}(2011)\citenamefont
  {Guillaume}, \citenamefont {Gregoryanz}, \citenamefont {Degtyareva},
  \citenamefont {McMahon}, \citenamefont {Hanfland}, \citenamefont {Evans},
  \citenamefont {Guthrie}, \citenamefont {Sinogeikin},\ and\ \citenamefont
  {Mao}}]{lithium-melt}%
  \BibitemOpen
  \bibfield  {author} {\bibinfo {author} {\bibfnamefont {C.~L.}\ \bibnamefont
  {Guillaume}}, \bibinfo {author} {\bibfnamefont {E.}~\bibnamefont
  {Gregoryanz}}, \bibinfo {author} {\bibfnamefont {O.}~\bibnamefont
  {Degtyareva}}, \bibinfo {author} {\bibfnamefont {M.~I.}\ \bibnamefont
  {McMahon}}, \bibinfo {author} {\bibfnamefont {M.}~\bibnamefont {Hanfland}},
  \bibinfo {author} {\bibfnamefont {S.}~\bibnamefont {Evans}}, \bibinfo
  {author} {\bibfnamefont {M.}~\bibnamefont {Guthrie}}, \bibinfo {author}
  {\bibfnamefont {S.~V.}\ \bibnamefont {Sinogeikin}}, \ and\ \bibinfo {author}
  {\bibfnamefont {H.}~\bibnamefont {Mao}},\ }\href@noop {} {\bibfield
  {journal} {\bibinfo  {journal} {Nature Physics}\ }\textbf {\bibinfo {volume}
  {7}},\ \bibinfo {pages} {211} (\bibinfo {year} {2011})}\BibitemShut {NoStop}%
\bibitem [{\citenamefont {Dahl}\ and\ \citenamefont
  {Stevenson}(2010)}]{turbulent-mixing-planetary-accretion}%
  \BibitemOpen
  \bibfield  {author} {\bibinfo {author} {\bibfnamefont {T.~W.}\ \bibnamefont
  {Dahl}}\ and\ \bibinfo {author} {\bibfnamefont {D.~J.}\ \bibnamefont
  {Stevenson}},\ }\href@noop {} {\bibfield  {journal} {\bibinfo  {journal}
  {Earth and Planetary Science Letters}\ }\textbf {\bibinfo {volume} {295}},\
  \bibinfo {pages} {177} (\bibinfo {year} {2010})}\BibitemShut {NoStop}%
\bibitem [{\citenamefont {Dewaele}\ \emph {et~al.}(2010)\citenamefont
  {Dewaele}, \citenamefont {Mezouar}, \citenamefont {Guignot},\ and\
  \citenamefont {Loubeyre}}]{dewale-ta-melt}%
  \BibitemOpen
  \bibfield  {author} {\bibinfo {author} {\bibfnamefont {A.}~\bibnamefont
  {Dewaele}}, \bibinfo {author} {\bibfnamefont {M.}~\bibnamefont {Mezouar}},
  \bibinfo {author} {\bibfnamefont {N.}~\bibnamefont {Guignot}}, \ and\
  \bibinfo {author} {\bibfnamefont {P.}~\bibnamefont {Loubeyre}},\ }\href
  {\doibase 10.1103/PhysRevLett.104.255701} {\bibfield  {journal} {\bibinfo
  {journal} {Phys. Rev. Lett.}\ }\textbf {\bibinfo {volume} {104}},\ \bibinfo
  {pages} {255701} (\bibinfo {year} {2010})}\BibitemShut {NoStop}%
\bibitem [{\citenamefont {Anzellini}\ \emph {et~al.}(2013)\citenamefont
  {Anzellini}, \citenamefont {Dewaele}, \citenamefont {Mezouar}, \citenamefont
  {Loubeyre},\ and\ \citenamefont {Morard}}]{dewale-Fe-melt}%
  \BibitemOpen
  \bibfield  {author} {\bibinfo {author} {\bibfnamefont {S.}~\bibnamefont
  {Anzellini}}, \bibinfo {author} {\bibfnamefont {A.}~\bibnamefont {Dewaele}},
  \bibinfo {author} {\bibfnamefont {M.}~\bibnamefont {Mezouar}}, \bibinfo
  {author} {\bibfnamefont {P.}~\bibnamefont {Loubeyre}}, \ and\ \bibinfo
  {author} {\bibfnamefont {G.}~\bibnamefont {Morard}},\ }\href@noop {}
  {\bibfield  {journal} {\bibinfo  {journal} {Science}\ }\textbf {\bibinfo
  {volume} {340}},\ \bibinfo {pages} {464} (\bibinfo {year}
  {2013})}\BibitemShut {NoStop}%
\bibitem [{\citenamefont {Ross}\ \emph {et~al.}(2005)\citenamefont {Ross},
  \citenamefont {Boehler},\ and\ \citenamefont {Soderlind}}]{dac-xe-melt}%
  \BibitemOpen
  \bibfield  {author} {\bibinfo {author} {\bibfnamefont {M.}~\bibnamefont
  {Ross}}, \bibinfo {author} {\bibfnamefont {R.}~\bibnamefont {Boehler}}, \
  and\ \bibinfo {author} {\bibfnamefont {P.}~\bibnamefont {Soderlind}},\ }\href
  {\doibase 10.1103/PhysRevLett.95.257801} {\bibfield  {journal} {\bibinfo
  {journal} {Phys. Rev. Lett.}\ }\textbf {\bibinfo {volume} {95}},\ \bibinfo
  {pages} {257801} (\bibinfo {year} {2005})}\BibitemShut {NoStop}%
\bibitem [{\citenamefont {Belonoshko}\ \emph {et~al.}(2006)\citenamefont
  {Belonoshko}, \citenamefont {Davis}, \citenamefont {Rosengren}, \citenamefont
  {Ahuja}, \citenamefont {Johansson}, \citenamefont {Simak}, \citenamefont
  {Burakovsky},\ and\ \citenamefont {Preston}}]{belonoshko-xe-melt}%
  \BibitemOpen
  \bibfield  {author} {\bibinfo {author} {\bibfnamefont {A.~B.}\ \bibnamefont
  {Belonoshko}}, \bibinfo {author} {\bibfnamefont {S.}~\bibnamefont {Davis}},
  \bibinfo {author} {\bibfnamefont {A.}~\bibnamefont {Rosengren}}, \bibinfo
  {author} {\bibfnamefont {R.}~\bibnamefont {Ahuja}}, \bibinfo {author}
  {\bibfnamefont {B.}~\bibnamefont {Johansson}}, \bibinfo {author}
  {\bibfnamefont {S.~I.}\ \bibnamefont {Simak}}, \bibinfo {author}
  {\bibfnamefont {L.}~\bibnamefont {Burakovsky}}, \ and\ \bibinfo {author}
  {\bibfnamefont {D.~L.}\ \bibnamefont {Preston}},\ }\href {\doibase
  10.1103/PhysRevB.74.054114} {\bibfield  {journal} {\bibinfo  {journal} {Phys.
  Rev. B}\ }\textbf {\bibinfo {volume} {74}},\ \bibinfo {pages} {054114}
  (\bibinfo {year} {2006})}\BibitemShut {NoStop}%
\bibitem [{\citenamefont {Armiento}\ and\ \citenamefont
  {Mattsson}(2005)}]{AM05}%
  \BibitemOpen
  \bibfield  {author} {\bibinfo {author} {\bibfnamefont {R.}~\bibnamefont
  {Armiento}}\ and\ \bibinfo {author} {\bibfnamefont {A.~E.}\ \bibnamefont
  {Mattsson}},\ }\href {\doibase 10.1103/PhysRevB.72.085108} {\bibfield
  {journal} {\bibinfo  {journal} {Phys. Rev. B}\ }\textbf {\bibinfo {volume}
  {72}},\ \bibinfo {pages} {085108} (\bibinfo {year} {2005})}\BibitemShut
  {NoStop}%
\bibitem [{\citenamefont {Andersson}\ \emph {et~al.}(1996)\citenamefont
  {Andersson}, \citenamefont {Langreth},\ and\ \citenamefont
  {Lundqvist}}]{vdw-dft}%
  \BibitemOpen
  \bibfield  {author} {\bibinfo {author} {\bibfnamefont {Y.}~\bibnamefont
  {Andersson}}, \bibinfo {author} {\bibfnamefont {D.~C.}\ \bibnamefont
  {Langreth}}, \ and\ \bibinfo {author} {\bibfnamefont {B.~I.}\ \bibnamefont
  {Lundqvist}},\ }\href {\doibase 10.1103/PhysRevLett.76.102} {\bibfield
  {journal} {\bibinfo  {journal} {Phys. Rev. Lett.}\ }\textbf {\bibinfo
  {volume} {76}},\ \bibinfo {pages} {102} (\bibinfo {year} {1996})}\BibitemShut
  {NoStop}%
\bibitem [{\citenamefont {Nellis}\ \emph {et~al.}(1982)\citenamefont {Nellis},
  \citenamefont {van Thiel},\ and\ \citenamefont {Mitchell}}]{nellis-shock-xe}%
  \BibitemOpen
  \bibfield  {author} {\bibinfo {author} {\bibfnamefont {W.~J.}\ \bibnamefont
  {Nellis}}, \bibinfo {author} {\bibfnamefont {M.}~\bibnamefont {van Thiel}}, \
  and\ \bibinfo {author} {\bibfnamefont {A.~C.}\ \bibnamefont {Mitchell}},\
  }\href {\doibase 10.1103/PhysRevLett.48.816} {\bibfield  {journal} {\bibinfo
  {journal} {Phys. Rev. Lett.}\ }\textbf {\bibinfo {volume} {48}},\ \bibinfo
  {pages} {816} (\bibinfo {year} {1982})}\BibitemShut {NoStop}%
\bibitem [{\citenamefont {Foulkes}\ \emph {et~al.}(2001)\citenamefont
  {Foulkes}, \citenamefont {Mitas}, \citenamefont {Needs},\ and\ \citenamefont
  {Rajagopal}}]{dmc-review}%
  \BibitemOpen
  \bibfield  {author} {\bibinfo {author} {\bibfnamefont {W.~M.~C.}\
  \bibnamefont {Foulkes}}, \bibinfo {author} {\bibfnamefont {L.}~\bibnamefont
  {Mitas}}, \bibinfo {author} {\bibfnamefont {R.~J.}\ \bibnamefont {Needs}}, \
  and\ \bibinfo {author} {\bibfnamefont {G.}~\bibnamefont {Rajagopal}},\ }\href
  {\doibase 10.1103/RevModPhys.73.33} {\bibfield  {journal} {\bibinfo
  {journal} {Rev. Mod. Phys.}\ }\textbf {\bibinfo {volume} {73}},\ \bibinfo
  {pages} {33} (\bibinfo {year} {2001})}\BibitemShut {NoStop}%
\bibitem [{\citenamefont {Drummond}\ and\ \citenamefont
  {Needs}(2006)}]{cambridge-dmc-noble-gas}%
  \BibitemOpen
  \bibfield  {author} {\bibinfo {author} {\bibfnamefont {N.~D.}\ \bibnamefont
  {Drummond}}\ and\ \bibinfo {author} {\bibfnamefont {R.~J.}\ \bibnamefont
  {Needs}},\ }\href {\doibase 10.1103/PhysRevB.73.024107} {\bibfield  {journal}
  {\bibinfo  {journal} {Phys. Rev. B}\ }\textbf {\bibinfo {volume} {73}},\
  \bibinfo {pages} {024107} (\bibinfo {year} {2006})}\BibitemShut {NoStop}%
\bibitem [{\citenamefont {Sorella}\ \emph {et~al.}(2007)\citenamefont
  {Sorella}, \citenamefont {Casula},\ and\ \citenamefont
  {Rocca}}]{casula-benzene-dimer}%
  \BibitemOpen
  \bibfield  {author} {\bibinfo {author} {\bibfnamefont {S.}~\bibnamefont
  {Sorella}}, \bibinfo {author} {\bibfnamefont {M.}~\bibnamefont {Casula}}, \
  and\ \bibinfo {author} {\bibfnamefont {D.}~\bibnamefont {Rocca}},\ }\href
  {\doibase DOI:10.1063/1.2746035} {\bibfield  {journal} {\bibinfo  {journal}
  {Journal of Chemical Physics}\ }\textbf {\bibinfo {volume} {127}},\ \bibinfo
  {pages} {014105} (\bibinfo {year} {2007})}\BibitemShut {NoStop}%
\bibitem [{\citenamefont {Beaudet}\ \emph {et~al.}(2008)\citenamefont
  {Beaudet}, \citenamefont {Casula}, \citenamefont {Kim}, \citenamefont
  {Sorella},\ and\ \citenamefont {Martin}}]{beaudet-h2-benzene}%
  \BibitemOpen
  \bibfield  {author} {\bibinfo {author} {\bibfnamefont {T.~D.}\ \bibnamefont
  {Beaudet}}, \bibinfo {author} {\bibfnamefont {M.}~\bibnamefont {Casula}},
  \bibinfo {author} {\bibfnamefont {J.}~\bibnamefont {Kim}}, \bibinfo {author}
  {\bibfnamefont {S.}~\bibnamefont {Sorella}}, \ and\ \bibinfo {author}
  {\bibfnamefont {R.~M.}\ \bibnamefont {Martin}},\ }\href {\doibase
  DOI:10.1063/1.2987716} {\bibfield  {journal} {\bibinfo  {journal} {Journal of
  Chemical Physics}\ }\textbf {\bibinfo {volume} {129}},\ \bibinfo {pages}
  {164711} (\bibinfo {year} {2008})}\BibitemShut {NoStop}%
\bibitem [{\citenamefont {Vinet}\ \emph {et~al.}(1986)\citenamefont {Vinet},
  \citenamefont {Ferrante}, \citenamefont {Smith},\ and\ \citenamefont
  {Rose}}]{vinet}%
  \BibitemOpen
  \bibfield  {author} {\bibinfo {author} {\bibfnamefont {P.}~\bibnamefont
  {Vinet}}, \bibinfo {author} {\bibfnamefont {J.}~\bibnamefont {Ferrante}},
  \bibinfo {author} {\bibfnamefont {J.~R.}\ \bibnamefont {Smith}}, \ and\
  \bibinfo {author} {\bibfnamefont {J.~H.}\ \bibnamefont {Rose}},\ }\href
  {http://stacks.iop.org/0022-3719/19/i=20/a=001} {\bibfield  {journal}
  {\bibinfo  {journal} {Journal of Physics C: Solid State Physics}\ }\textbf
  {\bibinfo {volume} {19}},\ \bibinfo {pages} {L467} (\bibinfo {year}
  {1986})}\BibitemShut {NoStop}%
\bibitem [{\citenamefont {Sola}\ and\ \citenamefont
  {Alf\`e}(2009)}]{sola-alfe-fe-melt}%
  \BibitemOpen
  \bibfield  {author} {\bibinfo {author} {\bibfnamefont {E.}~\bibnamefont
  {Sola}}\ and\ \bibinfo {author} {\bibfnamefont {D.}~\bibnamefont {Alf\`e}},\
  }\href {\doibase 10.1103/PhysRevLett.103.078501} {\bibfield  {journal}
  {\bibinfo  {journal} {Phys. Rev. Lett.}\ }\textbf {\bibinfo {volume} {103}},\
  \bibinfo {pages} {078501} (\bibinfo {year} {2009})}\BibitemShut {NoStop}%
\bibitem [{\citenamefont {Boehler}(1993)}]{dac-melt-fe}%
  \BibitemOpen
  \bibfield  {author} {\bibinfo {author} {\bibfnamefont {R.}~\bibnamefont
  {Boehler}},\ }\href {\doibase 10.1038/363534a0} {\bibfield  {journal}
  {\bibinfo  {journal} {Nature}\ }\textbf {\bibinfo {volume} {363}},\ \bibinfo
  {pages} {534 } (\bibinfo {year} {1993})}\BibitemShut {NoStop}%
\bibitem [{\citenamefont {Bouchet}\ \emph {et~al.}(2009)\citenamefont
  {Bouchet}, \citenamefont {Bottin}, \citenamefont {Jomard},\ and\
  \citenamefont {Z\'erah}}]{al-melt-paper}%
  \BibitemOpen
  \bibfield  {author} {\bibinfo {author} {\bibfnamefont {J.}~\bibnamefont
  {Bouchet}}, \bibinfo {author} {\bibfnamefont {F.}~\bibnamefont {Bottin}},
  \bibinfo {author} {\bibfnamefont {G.}~\bibnamefont {Jomard}}, \ and\ \bibinfo
  {author} {\bibfnamefont {G.}~\bibnamefont {Z\'erah}},\ }\href {\doibase
  10.1103/PhysRevB.80.094102} {\bibfield  {journal} {\bibinfo  {journal} {Phys.
  Rev. B}\ }\textbf {\bibinfo {volume} {80}},\ \bibinfo {pages} {094102}
  (\bibinfo {year} {2009})}\BibitemShut {NoStop}%
\bibitem [{\citenamefont {Root}\ \emph {et~al.}(2010)\citenamefont {Root},
  \citenamefont {Magyar}, \citenamefont {Carpenter}, \citenamefont {Hanson},\
  and\ \citenamefont {Mattsson}}]{root-xe-hugoniot}%
  \BibitemOpen
  \bibfield  {author} {\bibinfo {author} {\bibfnamefont {S.}~\bibnamefont
  {Root}}, \bibinfo {author} {\bibfnamefont {R.~J.}\ \bibnamefont {Magyar}},
  \bibinfo {author} {\bibfnamefont {J.~H.}\ \bibnamefont {Carpenter}}, \bibinfo
  {author} {\bibfnamefont {D.~L.}\ \bibnamefont {Hanson}}, \ and\ \bibinfo
  {author} {\bibfnamefont {T.~R.}\ \bibnamefont {Mattsson}},\ }\href {\doibase
  10.1103/PhysRevLett.105.085501} {\bibfield  {journal} {\bibinfo  {journal}
  {Phys. Rev. Lett.}\ }\textbf {\bibinfo {volume} {105}},\ \bibinfo {pages}
  {085501} (\bibinfo {year} {2010})}\BibitemShut {NoStop}%
\bibitem [{\citenamefont {Kresse}\ and\ \citenamefont
  {Furthmuller}(1996)}]{vasp}%
  \BibitemOpen
  \bibfield  {author} {\bibinfo {author} {\bibfnamefont {G.}~\bibnamefont
  {Kresse}}\ and\ \bibinfo {author} {\bibfnamefont {J.}~\bibnamefont
  {Furthmuller}},\ }\href {\doibase 10.1103/PhysRevB.54.11169} {\bibfield
  {journal} {\bibinfo  {journal} {Phys. Rev. B}\ }\textbf {\bibinfo {volume}
  {54}},\ \bibinfo {pages} {11169} (\bibinfo {year} {1996})}\BibitemShut
  {NoStop}%
\bibitem [{\citenamefont {Saija}\ and\ \citenamefont
  {Prestipino}(2005)}]{saija}%
  \BibitemOpen
  \bibfield  {author} {\bibinfo {author} {\bibfnamefont {F.}~\bibnamefont
  {Saija}}\ and\ \bibinfo {author} {\bibfnamefont {S.}~\bibnamefont
  {Prestipino}},\ }\href@noop {} {\bibfield  {journal} {\bibinfo  {journal}
  {Phys. Rev. B}\ }\textbf {\bibinfo {volume} {72}},\ \bibinfo {pages} {024113}
  (\bibinfo {year} {2005})}\BibitemShut {NoStop}%
\bibitem [{\citenamefont {Boehler}\ \emph {et~al.}(2001)\citenamefont
  {Boehler}, \citenamefont {Ross}, \citenamefont {S\"oderlind},\ and\
  \citenamefont {Boercker}}]{boehler-xe-melt}%
  \BibitemOpen
  \bibfield  {author} {\bibinfo {author} {\bibfnamefont {R.}~\bibnamefont
  {Boehler}}, \bibinfo {author} {\bibfnamefont {M.}~\bibnamefont {Ross}},
  \bibinfo {author} {\bibfnamefont {P.}~\bibnamefont {S\"oderlind}}, \ and\
  \bibinfo {author} {\bibfnamefont {D.~B.}\ \bibnamefont {Boercker}},\
  }\href@noop {} {\bibfield  {journal} {\bibinfo  {journal} {Phys. Rev. Lett.}\
  }\textbf {\bibinfo {volume} {86}},\ \bibinfo {pages} {5731} (\bibinfo {year}
  {2001})}\BibitemShut {NoStop}%
\bibitem [{\citenamefont {Kechin}(2001)}]{ketchin-melt}%
  \BibitemOpen
  \bibfield  {author} {\bibinfo {author} {\bibfnamefont {V.~V.}~\bibnamefont
  {Kechin}},\ }\href@noop {} {\bibfield  {journal} {\bibinfo  {journal} {Phys.
  Rev. B}\ }\textbf {\bibinfo {volume} {65}},\ \bibinfo {pages} {052102}
  (\bibinfo {year} {2001})}\BibitemShut {NoStop}%
\bibitem [{\citenamefont {Belonoshko}\ and\ \citenamefont
  {Dubrovinsky}(1997)}]{explain-melt-anomalies}%
  \BibitemOpen
  \bibfield  {author} {\bibinfo {author} {\bibfnamefont {A.~B.}\ \bibnamefont
  {Belonoshko}}\ and\ \bibinfo {author} {\bibfnamefont {L.~S.}\ \bibnamefont
  {Dubrovinsky}},\ }\href@noop {} {\bibfield  {journal} {\bibinfo  {journal}
  {American Mineralogist}\ }\textbf {\bibinfo {volume} {82}},\ \bibinfo {pages}
  {441} (\bibinfo {year} {1997})}\BibitemShut {NoStop}%
\end{thebibliography}%

\end{document}